\documentclass[final,3p,times,twocolumn]{elsarticle}

\usepackage{graphicx}
\usepackage{epsfig}

\usepackage{amssymb}

\journal{Acta Meterialia}

\begin{document}

\begin{frontmatter}



\title{ Hidden order in serrated flow of metallic glasses }

\author[label1]{Ritupan Sarmah}
\ead{ritupan@mrc.iisc.ernet.in}
\author[label1]{ G. Ananthakrishna \corref{cor1} }
\ead{garani@mrc.iisc.ernet.in}
\cortext[cor1]{Corresponding author. Tel.: +91 80 22932780}
\author[label2]{B.A. Sun}
\author[label2]{W. H. Wang}

\address[label1]{Materials Research Centre, Indian Institute of Science, Bangalore 560012, India }

\address[label2]{Institute of Physics, Chinese Academy of Sciences, Beijing 100190, P. R. China}

\begin{abstract}

We report results of statistical and dynamic analysis of the serrated stress-time curves obtained from  compressive  constant strain-rate  tests on two metallic glass samples with different ductility levels in an effort to extract hidden  information in the seemingly irregular serrations.  Two distinct types of dynamics are detected in  these two alloy samples.  The stress-strain curve corresponding to the less ductile $Zr_{65}Cu_{15}Ni_{10}Al_{10}$  alloy is shown to  exhibit finite correlation dimension and a positive Lyapunov exponent, suggesting that the underlying dynamics is chaotic. In contrast, for the more ductile $Cu_{47.5}Zr_{47.5}Al_{5}$ alloy, the distributions of stress drop magnitudes and their  time durations obey  a power law scaling reminiscent of a self-organized critical state. The exponents also satisfy the scaling  relation compatible with self-organized criticality. Possible physical mechanisms contributing to the two distinct dynamic regimes are discussed by drawing on the analogy with the serrated yielding of crystalline samples. The analysis, together with some physical reasoning, suggests that plasticity in the less ductile sample can be attributed to stick-slip of single shear band, while that of the more ductile sample could be attributed to the simultaneous nucleation of large number of shear bands and their mutual interactions.

\end{abstract}

\begin{keyword}
Metallic glasses \sep Ductility \sep  Intermittent flow \sep Chaos \sep Self-organized criticality

\end{keyword}

\end{frontmatter}



\section{Introduction}

The discontinuous yielding of crystalline metallic alloys during plastic deformation is well known to exhibit temporal fluctuations 
in the stress-strain curves. These fluctuations are accompanied by spatial localization of dislocations into bands \cite {Cottrell}. This phenomenon, known as the Portevin-Le Chatelier (PLC) effect \cite{PLC}, has been studied extensively in crystalline materials, and 
continues to attract considerable attention \cite{GA07,Anan04,Kok03,Bhar03a,Hahner02,Bhar01,Bhar02a,GA99}. The physical origin of the intermittent flow is attributed to the collective behavior of dislocations \cite{Cottrell,GA07}. In contrast, the less-well known, but similar serrated flow along with localization of deformation into shear bands has been observed in a large number of studies on bulk metallic glasses (BMGs) despite the fact that BMGs are characterized by the absence of long-range order. Although dislocation-mediated deformation mechanism is absent in BMGs, serrated flow phenomenon in the plastic regime has been widely observed in nanoindentation experiments \cite{Nieh03,Yang07,Jiang03,Greer04,Jang07} and  compression tests \cite{Johnson04,Greer06,Wright,Jiang08,Kumar07,Yu08}. The plastic deformation of BMGs at room temperature is characterized by nanoscale shear bands that contribute to strain softening arising from dilation \cite {Spaepen77}, or an increase in temperature, or both \cite {Greer06,Wang72}. These serrations have been attributed to shear-banding operation. As the strain rate increases, the amplitude of serrations decreases with fewer shear bands simultaneously contributing to serrations at high strain rates compared to their number at low strain rates \cite{Jiang08}. In the case of nanoindentations, the serrations have been reported to disappear at high rates of indentation. This has been attributed to the inability of a single shear band to accommodate the entire shear at high rates \cite{Nieh03}.  However, Jiang and Atzmon \cite{Jiang03} and Greer {\it et al.} \cite{Greer04} have considered the disappearance of serrations at high strain rates as arising from limitations of data acquisition or due to proliferation of shear bands. Subsequent studies demonstrated that the deformation remains inhomogeneous even at high indentation rates \cite{Jang07}, which is also consistent with results on compression tests \cite{Jiang08}.

The inhomogeneous deformation of metallic glasses has also been studied theoretically.  The conventional picture of the deformation of 
amorphous materials is based on the free volume model \cite{Spaepen77,Spaepen82} or the shear transformation zones (STZ) model \cite{Argon79,Falk98}, which ascribe the macroscopic deformation to the presence of localized deformation centers. This is also supported by numerical simulations \cite{Falk98,Lund03N,Zink06,Argon05,Bailey07,Haxton07}.  Such shear transformations should lead to long-range interaction. Models that include such interactions do exhibit localized deformation patterns that resemble the shear bands \cite{Argon05,Langer01,Langer06}. Another direction of investigation of intermittent plastic flow recognizes the importance of nonlinearities that naturally lead to loss of stability of the ''elastic branch'' or equivalently to loss of analyticity of stress \cite{karmakar10}.

Here we adopt a completely new approach to analyze the serrated flow of BMGs in an effort to obtain insight in the deformation dynamics of these materials. In general,  the serrated stress-strain curves of BMGs appear quite complex and irregular. The question we address in this paper is whether the seemingly irregular nature of  stress-strain curves  is purely random or there is any hidden order that could extracted from them and what physical conclusions could be drawn from such a study.  Indeed, such efforts have been carried out on serrated flow of the crystalline alloys, namely the PLC effect \cite{Bhar01,Bhar02a,GA99,Noro97}. The methodology rooted in the area of nonlinear dynamics, called nonlinear time series analysis, is designed  to {\it recover hidden information from the seemingly irregular stress-strain series alone.}  The well-developed subject of  nonlinear dynamics suggests that any irregular time series could well arise due to nonlinear deterministic interaction of at least three degrees of freedom, referred to as{\it chaos}, instead of being purely random in origin. Indeed, such irregular time series are found in a large number of physical situations including physical, chemical and biological systems which have been demonstrated to be chaotic \cite{KS97}.

That such an approach could be useful is also suggested by the fact that the intermittent flow of BMGs has been recognized to be a 
stick-slip phenomenon \cite{Cheng09,Song08}. Indeed, a large number of systems ranging from atomic scale (e.g. atomic force microscopy) 
to the geological scale (earthquakes) exhibit stick-slip dynamics. At laboratory scale, stick-slip phenomenon has been observed in a large number of rate-controlled experiments such as the PLC effect \cite{GA07,Anan04,Kubin87}, peeling of an adhesive tape \cite{Jag08a,Jag08b}, micellar vesicles subjected to shear \cite{Sood06}, etc. The phenomenon is characterized by a sticking phase that usually lasts much longer than the slipping phase, a feature also observed in the serrations of BMG deformation.  Stick-slip dynamics is well established to result from a competition between the internal relaxation time scales and the time scale associated with drive rate \cite{Strogatz}. Indeed,  this is consistent with the fact that several time scales corresponding to free volume, temperature, etc., are well known to influence BMG deformation \cite{Spaepen77,Argon79}. While the above facts suggest that such an analysis is worthwhile,  further support also comes from the recently established similarity between the serrated flow in BMG deformation and the PLC effect for  crystalline 
alloys \cite{Torre07}. The latter is a  well established  example of stick-slip dynamics arising from the collective pinning and 
unpinning of dislocations from solute atmosphere \cite{Cottrell,GA07,Anan04}.

In view of the similarities between the serrated flow in BMGs and the PLC effect, we recall some relevant methods adopted in 
analyzing the serrated flow of the PLC effect. Time-series analysis of the stress-strain curves from the PLC effect of single crystals 
and polycrystals have been carried out. Such an analysis was motivated by the prediction by the Ananthakrishna model 
\cite{GA07,Anan04,Bhar03a,Anan81,Anan82,Anan83} that the stress-strain curves can be chaotic for a range of strain rates.\footnote{Indeed, 
the Ananthakrishna model uses a dynamic approach to study the PLC effect \cite{Anan81,Anan82,Anan83} and considers the 
phenomenon as arising from nonlinear interaction of a few collective modes of dislocations \cite{Anan82}. The model recovers most 
generic features of the PLC effect apart from predicting the serrations to be chaotic \cite{GA07,Anan04,Bhar03a,Anan81,Anan82,Anan83}.}
The results of the analysis show the existence of two distinct types of dynamic regimes, namely chaotic and self-organized criticality
\cite{Bhar01,Bhar02a,GA99,Noro97}, in two different regimes of strain rate, consistent with the model prediction 
\cite{GA07,Anan04,Bhar03a,Bhar01,Bhar02a,GA99,Anan81,Anan82,Anan83}. {\it So far such a systematic dynamic analysis of the serrated 
flow of the BMGs has not been performed. It is the purpose of this paper to carry out dynamic time-series analysis of the stress-strain 
curves of two different types of BMGs.} We closely follow the methods used in Refs. \cite{Bhar01,Bhar02a,GA99,Noro97}. Simultaneously, 
we have carried out a detailed statistical analysis of stress drop magnitudes and durations \cite{Wang09}.  Recently Wang {\it et al.,} 
showed that  the statistics of the elastic energy density of several BMGs \cite{Wang09}  could be fitted to a power-law distribution  
modulated by a Gaussian function, suggesting that the BMG is in a state of self-organized criticality (SOC) \cite{Bak88,Bak96,Jensen98}. 
However, Wang {\it et al.,} did not carry out a full analysis involving the distributions of the durations of the events, as well as the consistency checks essential to assert SOC dynamics.  Our detailed analysis shows that while the less ductile $Zr_{65}Cu_{15}Ni_{10}Al_{10}$ BMG is chaotic, the relatively more ductile sample $Cu_{47.5}Zr_{47.5}Al_{5}$ surprisingly exhibits a power-law state of stress drops, suggesting self-organized critical state. Physical mechanisms contributing to these two distinct dynamic regimes are discussed. Our analysis also suggests that the high ductility of $Cu_{47.5}Zr_{47.5}Al_{5}$  sample emerges from the production of several shear bands and their interactions rather than partial plastic relaxation of the shear activity as in the PLC effect \cite{Anan04,Bhar03a,Bhar01,Bhar02a,GA99}.

\section{Experimental procedure} 

$Zr_{65}Cu_{15}Ni_{10}Al_{10}$ and $Cu_{47.5}Zr_{47.5}Al_{5}$ alloys with nominal compositions were prepared by arc melting a mixture of pure  metal elements in an argon atmosphere, followed by suction casting into copper molds to form rod-like BMG samples with a diameter of $2$mm. The glassy phase of the as-cast BMG specimens were examined  by transmission electron microscope (TEM) and  X-ray diffraction (XRD).  The specimen surface was examined by scanning electron microscope (SEM; Philips XL $30$). The compressive test specimens were cut from the BMG rods with a nominal  2:1 aspect ratio. The samples were carefully ground to $1\mu m$ accuracy. Compressive tests were performed in a Instron electromechanical $3384$ test system at a constant strain rate of $1\times 10^{-4} s^{-1}$. Care was exercised to ensure that the two surfaces were nearly parallel and orthogonal to the loading axis.  Load, displacement and time were measured and digitally stored at a frequency of $50$ Hz ($50$ datum point per second) for the $Zr_{65}Cu_{15}Ni_{10}Al_{10}$ 
BMG sample and $20 Hz$ for the $Cu_{47.5}Zr_{47.5}Al_{5}$ BMG sample.

\begin{figure}[h]
\centering
\vbox{
\includegraphics[height=5.0cm,width=8.25cm]{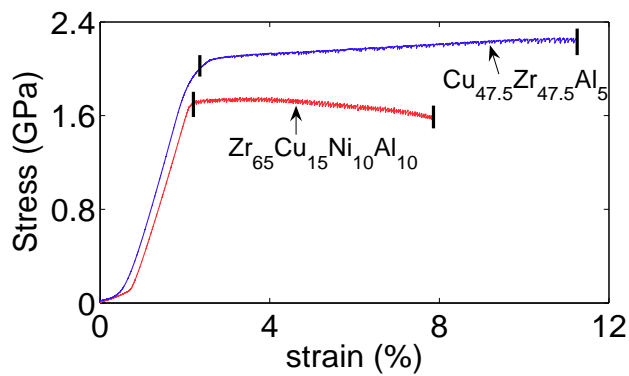}
}
\caption{Stress-strain curves for the two BMGs of composition $Zr_{65}Cu_{15}Ni_{10}Al_{10}$ and $Cu_{47.5}Zr_{47.5} Al_{5}$, respectively. 
}
\label{OrigSScurve}
\end{figure}

\begin{figure}
\centering
\includegraphics[height=5.0cm,width=8.5cm]{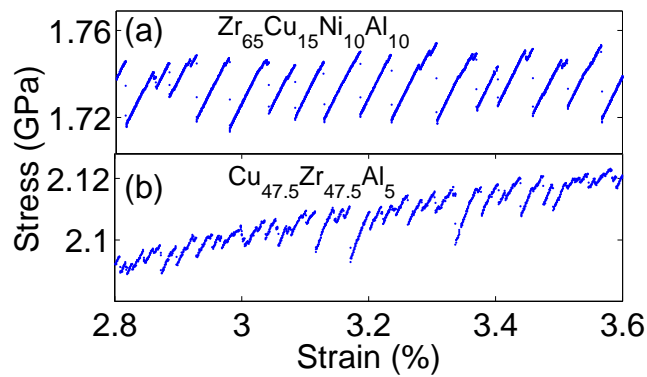}
\caption{(a,b) A small portion of the stress-strain curves  for the $Zr_{65}Cu_{15}Ni_{10}Al_{10}$ sample and $Cu_{47.5}Zr_{47.5} Al_{5}$ sample, respectively. }
\label{OrigSScurve1}
\end{figure}
\begin{figure}[h]
\center
\includegraphics[height=10.0cm,width=6.5cm]{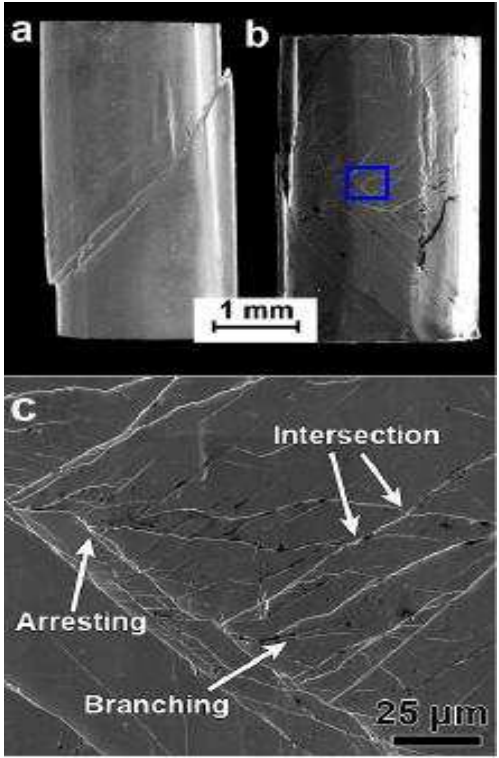}
\caption{(a and b) Shear bands for the less ductile  $Zr_{65}Cu_{15}Ni_{10}Al_{10}$ BMG sample and the more 
ductile $Cu_{47.5}Zr_{47.5} Al_{5}$ sample.  (c) Magnified view of the more ductile metallic glass where 
intersection, arrests and branching of the shear bands are shown. }
\label{Bands}
\end{figure}

Figure \ref{OrigSScurve} shows stress-strain curves for the two BMG samples. The lower curve corresponds to the sample of composition 
$Zr_{65}Cu_{15}Ni_{10}Al_{10}$  and the upper curve corresponds to the sample  of composition   $Cu_{47.5}Zr_{47.5}Al_{5}$. It is clear 
from the figure that the two alloys display significantly different plastic strains (about $8\%$ for $Zr_{65}Cu_{15}Ni_{10}Al_{10}$ 
and $11\%$ for $Cu_{47.5}Zr_{47.5}Al_{5}$) before the catastrophic failure. Thus, for convenience on notation, we label the less ductile $Zr_{65}Cu_{15}Ni_{10}Al_{10}$ BMG sample by $S_B$ and we shall generally refer to the more ductile $Cu_{47.5}Zr_{47.5}Al_{5}$ 
BMG sample $S_D$. Clearly, both samples exhibit intermittent flow after yielding at about $2\%$ elastic strain. The serrations are characterized by the repetitive cycles of abrupt stress drops followed by elastic reloading  (Fig. \ref{OrigSScurve1}a and b).  As can be seen, the serrations for   the $S_B$ ($Zr_{65}Cu_{15}Ni_{10}Al_{10}$) sample have large stress drops with very few small ones, while 
the serrations of the more ductile sample $S_D$ exhibits numerous small stress drops with fewer large  ones. Plots of stress-strain curves 
of the two BMGs for the same strain range are shown in Figure \ref{OrigSScurve1}. We note that sampling rate is such that for both samples, a few points are recorded on the region of stress drops.

Figure \ref{Bands} shows the shear band patterns on the surface of the deformed samples for the two BMGs. As shown in  Fig. \ref{Bands}a, a dominant shear band was observed during the deformation of the less ductile $Zr_{65}Cu_{15}Ni_{10}Al_{10}$ BMG sample, even though a fair amount of ductility is exhibited by the sample.  The reasonably large plasticity when a single dominant shear band operates has been recently attributed to stick-slip process \cite{Cheng09,Song08,Han09}. In contrast, multiple shear bands were seen on the surface of the deformed sample for the more ductile BMG $Cu_{47.5}Zr_{47.5} Al_{5}$  as shown in Fig. \ref{Bands}b.  A magnified image that displays numerous shear bands when the sample is deformed significantly is shown in Fig. \ref{Bands}c. The simultaneous nucleation of large number of shear bands is responsible for the good plasticity of the ductile sample $Cu_{47.5}Zr_{47.5}Al_{5}$ \cite{Liu07}. (See section 5 for further discussion.) 

\begin{figure}
\vbox{
\includegraphics[height=4.5cm,width=8cm]{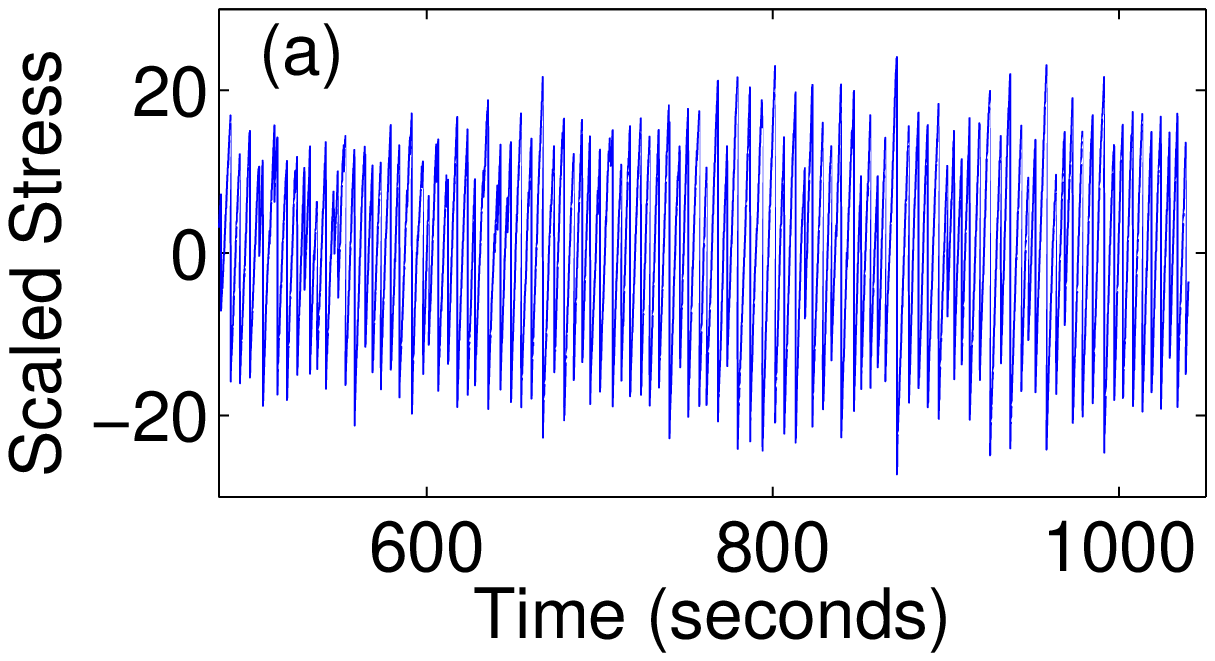}
\includegraphics[height=4.5cm,width=8cm]{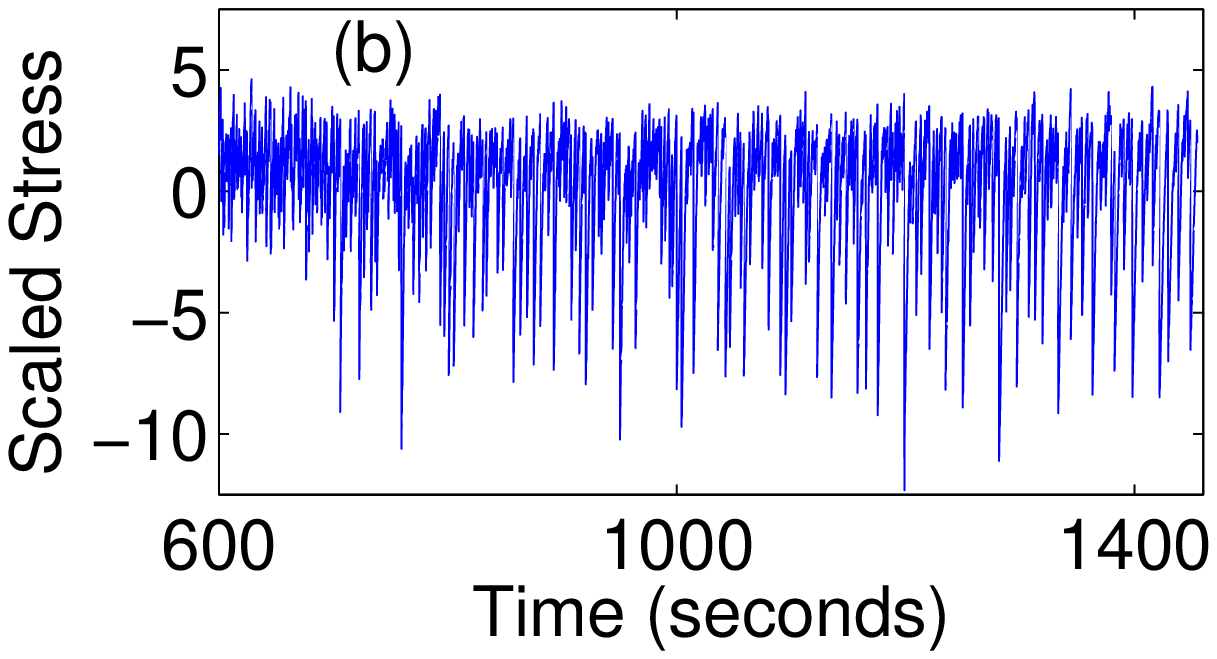}
}
\caption{(a and b)  Cleaned stress-time series for the less ductile $Zr_{65}Cu_{15}Ni_{10}Al_{10}$ BMG and the more ductile $Cu_{47.5}Zr_{47.5}Al_{5}$ BMG samples, respectively, for the region indicated in the original stress-strain plot. }
\label{CleanData}
\end{figure}

In general, work or strain hardening is not observed during the deformation of the BMGs as shear bands act as  work-softening sources attributed to shear dilation or temperature rise. In the present case, however, a clear upward drift in the stress-strain curve of the ductile sample $S_D$ can be observed. This ''work hardening'' behavior is expected to originate from the extensive multiplication of shear bands and consequent interaction between them \cite{Liu07}. Much like hardening in the PLC effect, the gradual increase in the stress level is also accompanied by increase in the magnitude of stress drops with increasing strain.

\section{Methods of analysis}

Given a time series, two kinds of analyses are possible. The first one is the statistical analysis of the time series. While the method is simple, it can yield substantial information on the nature of the underlying process. The second method is the more sophisticated method of  nonlinear time-series analysis. This approach is particularly useful if one suspects that the underlying process is of deterministic origin. As stick-slip process is a deterministic nonlinear phenomenon, the method of nonlinear time-series analysis  offers a platform for analyzing the phenomenon and also to understand the complexity of the stick-slip process. It will also help to establish if the irregular time series is chaotic. Both approaches assume infinitely long time series. We shall comment on the applicability of the methods to real systems.

\subsection{Statistical analysis: Power law distributions}

The simplest statistical quantity to compute is the statistics of events. The definition of an event depends on the physical situation, which in our case may be taken to be the magnitude of the stress drop. Since the stress drop  of a certain magnitude occurs in a certain time interval, the duration of the stress drop is also a characteristic feature of the event. In our analysis,  we have taken the difference between the maximum and next minimum to be a measure of the magnitude of the stress drop  denoted by $\Delta \sigma$.   Similarly, we define the time interval between a stress minimum to the previous maximum to be the duration of the stress drop denoted by $\Delta t$. It is clear that the corresponding distributions given by $D(\Delta \sigma)$ and $D(\Delta t)$ can be easily computed. As is the case for the PLC effect \cite{Bhar01,Bhar02a,GA99},  the distribution of the event sizes $D(\Delta \sigma)$ often follows a power-law scaling of the form
\begin{equation}
D(\Delta \sigma) \sim \Delta \sigma^{-\alpha},
\label{magnitude}
\end{equation}
where $\alpha$ is an exponent.   Similarly, $D(\Delta t)$ also follows a power law scaling of the form 
\begin{equation}
D(\Delta t) \sim \Delta t^{-\beta},
\label{magnitudeT}
\end{equation}
with $\beta$ being the corresponding exponent.  However, since $\Delta \sigma$ refers to the stress drop occurring in a time interval $\Delta t$, these two variables are not independent. Indeed, the basic statistical quantity describing the process is the joint probability of an event of magnitude $\Delta \sigma$ occurring in a time interval $\Delta t$ defined by $P(\Delta \sigma,\Delta t)$ \cite{Vives}.  Thus, both  $P(\Delta \sigma)$ and $P(\Delta t)$  are the marginal distributions obtained by integrating out the complimentary variable. (Note our $D's$ are unnormalized.) Thus, one can introduce another scaling exponent:
\begin{equation}
<\Delta t> \sim \Delta\sigma^x
\label{Tevent}
\end{equation}
where  $<.>$ is the conditional average. Then, the three exponents are connected through a scaling relation:
\begin{equation}
\alpha = x (\beta -1) + 1. 
\label{Srelation}
\end{equation}

\subsection{Time series analysis}

Quite often, only a scalar signal is measurable, which in our case is the stress. Such  signals can be   very irregular. This again is the case in our deformation experiments. Traditionally, such irregularity is  generally attributed to statistical noise such as the Brownian noise which can only be characterized by their distribution (or their moments). However, the well-developed methods in dynamic systems suggest that such  signals could well arise from nonlinear interaction of a few degrees freedom (at least three) called deterministic chaos or simply chaos. Two characteristic features of  chaos are the existence of a strange attractor with self-similar properties quantified by a fractal dimension (or equivalently the correlation dimension), and sensitivity to initial conditions quantified by the existence of a positive Lyapunov exponent.  Given the equations of motion, these quantities can be directly calculated. In contrast to the 
statistical analysis, time-series analysis  is a basic tool for establishing if a time series is of dynamical origin or not. This method is useful when a scalar time series is suspected to be a projection from a higher dimensional dynamics. Such time series are traditionally analyzed using embedding methods that attempt to recover the underlying dynamics.  The basic idea is to unfold the dynamics through a phase space reconstruction of the attractor by embedding the time series in a higher-dimensional space using a suitable time delay \cite{Packard80,GP83}. Consider a scalar time series measured in units of sampling time $\delta t$  defined by $[x(k),k=1,2,3,\cdots ,N]$. Then, one can construct $d-$dimensional vectors defined  by:   
\begin{eqnarray}
\nonumber 
\vec{\xi}_{k} &= &[x(k),x(k+\tau),\cdots,x(k+(d-1)\tau)],\\ 
 k&=&1,\cdots ,[N-(d-1)\tau].
\end{eqnarray}
The delay time $\tau$ suitable for the purpose is either obtained from the autocorrelation function or from mutual information \cite{KS97}. Once the reconstructed attractor is obtained, the existence of converged values of finite correlation dimension and a positive exponent is taken to be a signature of  the underlying chaotic dynamics.

\subsection{Correlation dimension}

The self-similar nature of the reconstructed attracted is quantified by using the popular algorithm due to Grassberger and Procaccia \cite{GP83} that calculates  the correlation integral. This is defined as the fraction of pairs of points $\vec{\xi}_{i}$ and $\vec{\xi}_{j}$ whose distance is  less than $r$ given by:
\begin{equation}
C(r)=\frac{1}{N_p}\sum_{i,j} \Theta(r-|\vec{\xi}_i-\vec{\xi}_j|),
\end{equation}
where $\Theta(\cdots)$ is the step function and $N_p$ the number of vector pairs summed. A window is imposed to exclude temporally correlated points \cite{KS97,Theiler86}.  The  method provides equivalence between the reconstructed attractor and the original  attractor. It has been shown that a proper equivalence is possible if the time series is noise free and long \cite{Ding93}. For a self similar attractor $C(r)\sim r^{\nu}$ in the limit of small $r$, where $\nu$ is the correlation dimension \cite{GP83}. Then, as $d$ is increased, one expects to find a convergence of the slopes $d \, ln \, C(r)/d \, ln \, r$ to a finite value in the limit of small $r$. However, in practice,  the scaling regime is found at  intermediate length scales due to limited length of the time series and the presence of noise.

\subsection{Lyapunov spectrum}

The existence of a positive Lyapunov exponent is considered as an unambiguous quantifier of chaotic dynamics. However, the presence of superposed noise component, which in the present case is high, poses problems.  In principle, the noise component can be cured and then the Lyapunov exponent calculated \cite{KS97}.  Here, we use an algorithm that does not require preprocessing of the data; it is designed to average out the influence of superposed noise. The algorithm, which is an extension of the Eckmann's algorithm, has been shown to work well for reasonably high levels of noise in model systems as well as for short time series (for details, see Ref. \cite{GA99,Noro01}). The method has also been used to analyze experimental time series. 

The conventional Eckmann's algorithm \cite{Eckmann86} relies on the construction of a sequence of tangent matrices $T_i$, which maps the difference vector $\vec{\xi}_i-\vec{\xi}_j$ to $\vec{\xi}_{i+k}-\vec{\xi}_{j+k}$ evolved for $k$ units of time, and successively reorthogonalize $T_i$ using standard $Q R$ decomposition ($Q$ is an orthogonal matrix and $R$ is an upper triangle matrix with positive diagonal elements). Then, the Lyapunov exponents are given by: 
\begin{equation}
\lambda_l = \frac{1}{kp\Delta t} \sum_{j=0}^{p-1} ln (R_j)_{ll}; \,\, l=1,2,\cdots d.
\label{Lyap}
\end{equation}
Here $p$ is the number of available matrices and $k$ is the propagation time in units of the sampling time $\delta t$. In the algorithm, the number of neighbors used is small, typically min$[2d,d+4]$, contained in a sphere of size $\epsilon_s$. A simple modification of this is to use those neighbors falling in shell $\epsilon_s$ defined by spheres of inner and outer radii $\epsilon_i$ and $\epsilon_0$, respectively. Then, ignoring points falling within the sphere, $\epsilon_i$ acts as a noise filter. However, these few neighbors will not be adequate to average out the noise  component superposed on the signal. Thus, the modification effected is to allow more neighbors so that the noise statistics is sampled properly. (For details, see Refs. \cite{GA99,Noro01}.) As  the sum of the exponents should be negative for a dissipative system, we impose this as a constraint. In addition, we also demand the existence of a stable positive exponent and a zero exponent (a necessary requirement for continuous time systems such as the stress-time series) over a finite range of shell 
sizes $\epsilon_s$. (Note that the constancy of the two exponents in the range of $\epsilon_s$ is equivalent to constancy with respect to the  number of neighbors contained in $\epsilon_s$ as it is varied over a certain range.)  The results presented below are for a specific value of the shell size that is in the stable range of shell sizes for which positive and zero Lyapunov exponents are constant.

Finally, the Lyapunov dimension can be calculated from the Lyapunov spectrum by using the Kaplan-Yorke conjecture, defined by the relation: 
\begin{equation} 
D_{KY} = j + \frac{\sum_1^{j}\lambda_{i}}{\vert \lambda_{j+1}\vert}, \,\, {\rm and } \,\, \sum_1^{j}\lambda_{i} > 0. 
\label{DKY}
\end{equation}
It has been conjectured that $D_{KY} \ge \nu$. For consistency, we have also calculated $D_{KY}$ \cite{KYorke}.

\section{Dynamics of serrated flow in bulk metallic glasses }

From a dynamic point of view, the irregular form of the stress-strain curves from the two BMG samples could be of deterministic nonlinear origin.  This point of view is also supported by the similarity of the stress-strain curves of the BMGs to  those based on the PLC effect,  coupled with  the fact that dynamic analysis of the PLC stress-strain curves have been shown to be chaotic \cite{Bhar01,Bhar02a,GA99,Noro97}.  Further  support comes from similar analysis of  other intermittent flows \cite{KS97,Jag08a,Jag08b,Sood06}.

\subsection{ Chaotic dynamics} 

A few comments are desirable on the limitations in  applying the above theoretical approaches developed for ideal systems to nonideal experimental stress-strain series. Nonlinear time-series analysis  assumes that the signal is of infinite extent and is stationary. In reality, however, experimental data are always of limited length.  More importantly, very often, the stationarity assumption is not satisfied, as is the case in the current problem. The first restriction is less serious and, indeed, algorithms have been developed to deal with short time series \cite{GA99,KS97,Noro01,Zeng91}. Such methods have been applied to a number of physical situations ranging from physics to biology \cite{KS97}. The second restriction is more serious. Non-stationarity is common in many real systems of practical relevance such as stock prices, intermittent natural phenomenon such as earthquakes, and signals from living matter. The specific type of non-stationarity relevant for the current study arises from the slow variation of internal state variables  compared to the rapidly fluctuating signal that is being analyzed \footnote{A good discussion on tests for stationarity in the context of nonlinear time-series analysis can be found in Ref. \cite{KS97}. The minimum requirement is that the total length of the time series should be much larger than some basic time scale, which in this case is the mean period of the stress fluctuations.}.  This is reflected in the systematic change in the mean stress level seen,  particularly,  in the ductile sample.  Nevertheless, if some insight about the underlying physical process is available, which is hardening in our case, this contribution can be subtracted to `'convert'' the non-stationary time series 
into a ''stationary one''. If the resulting time series is stationary, various statistical and dynamic quantities, such as the mean, variance, more importantly the correlation functions and correlation integrals should be the same (within some reasonable errors) for reasonably long subintervals of the whole time series.   Furthermore, in real systems, most experimental signals contain superposed noise, which in this case is high. There are several methods designed to cure the noise component including singular value decomposition \cite{KS97,BK86}. Usually, the cured data sets are then subjected to further analysis.

Apart from noise, there are three types of systematic variations in the time series. The first one corresponds to near-periodic fluctuations in the stress that can be traced to the vibrational frequency of the testing machine. The second is the accuracy in stress measurement, which is less than $0.5MPa$. The first error is easily eliminated. However, the accuracy of measurement is a limitation that is reflected in the extent of scaling regime in the correlation integral. The third is intrinsic to the material, namely the gradual increase (for the ductile  sample) in the mean stress level or the decrease in the mean stress level towards the end of the time series. The gradual increase in stress level is known to arise due to hardening and is considered as not relevant to the mechanisms 
controlling the stress drops.  Therefore, this component is eliminated by subtracting a running average from the original time series. As for the less ductile BMG sample $S_B$, the changes in the mean stress level, though small, are also dealt with in a similar manner. The stress drop magnitudes are also normalized by using a running average of the stress drops \cite{Bhar01,Bhar02a,GA99,Noro97}.  Further, the sample eventually breaks, and thus, not only the time series is limited, but the eventual failure of the sample can induce artefacts such as a decrease in the mean stress level. This also implies that we need to discard a certain portion of the time series (typically last 10-20 peaks) towards the end of the data set. (It may be noted that in relation to nonstationary effects, the artefacts 
arising from the sample breaking leading to limited  length of the time series is no different to the limited extent of time series due to other constraints as long as the data points closer to the break point are ignored.)  The resulting stress-strain series for both the samples shown in Fig. \ref{CleanData}a and b are used  for further analysis.

We first investigate the dynamics of the less ductile $Zr_{65}Cu_{15}Ni_{10}Al_{10}$  BMG. The total length of the time series used is about $29,000$ points in units of $\delta t=0.02 s$. Before proceeding further,  we first summarize the results of tests for stationarity (of the corrected time series), which involve calculating the mean, variance and autocorrelation function for several nonoverlapping subintervals of relatively long intervals.  As the original time series is of finite length, the calculated quantities should match within some reasonable errors.  We find these statistical quantities  agree within a small error (for subintervals of length of about $5000-7000$ points).  Using only two intervals, each containing about $14500$ points, we have also calculated the autocorrelation function  for the first and second half of the time series of the data set.  The calculated autocorrelation functions for the two halves were found to be nearly the same giving the same correlation time of $30$ units as for the full data set (and therefore not shown).

\begin{figure}[h]
\vbox{
\includegraphics[height=5cm,width=8cm]{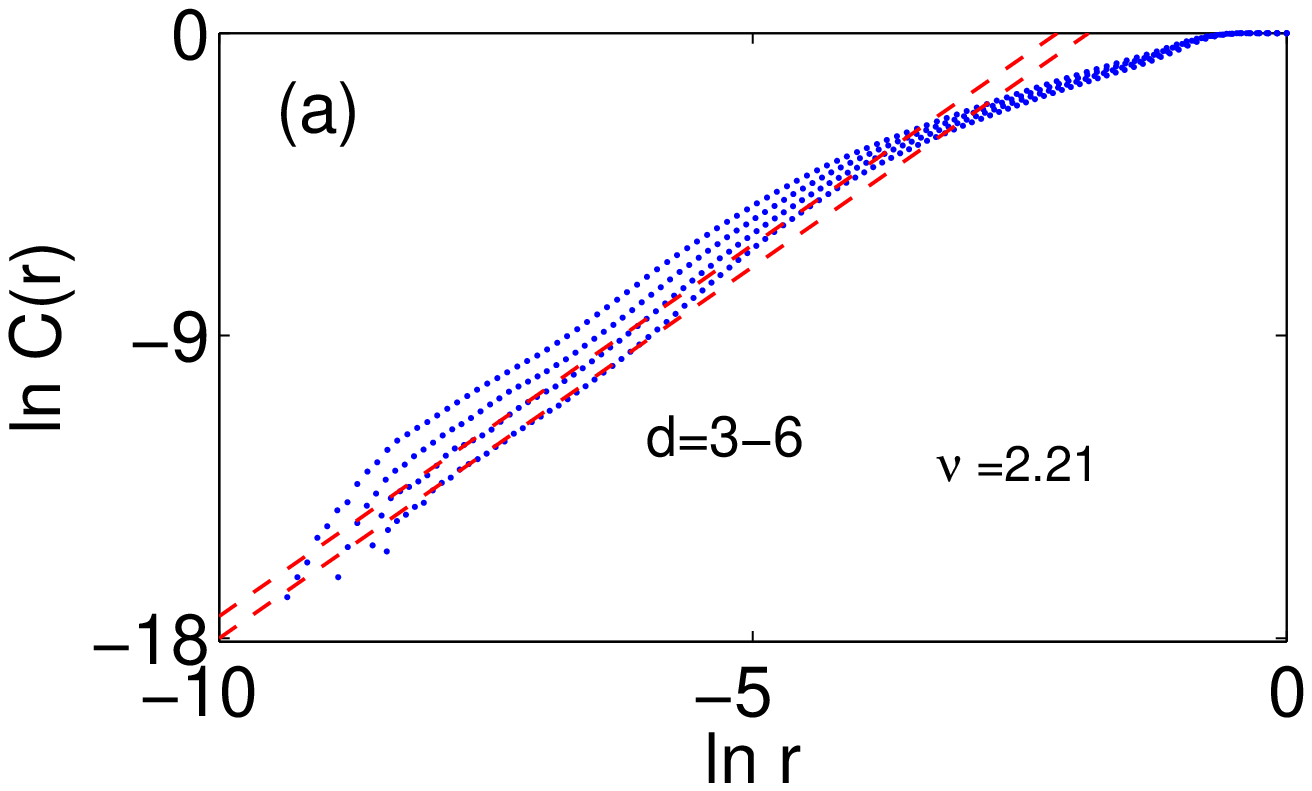}
\includegraphics[height=5cm,width=8cm]{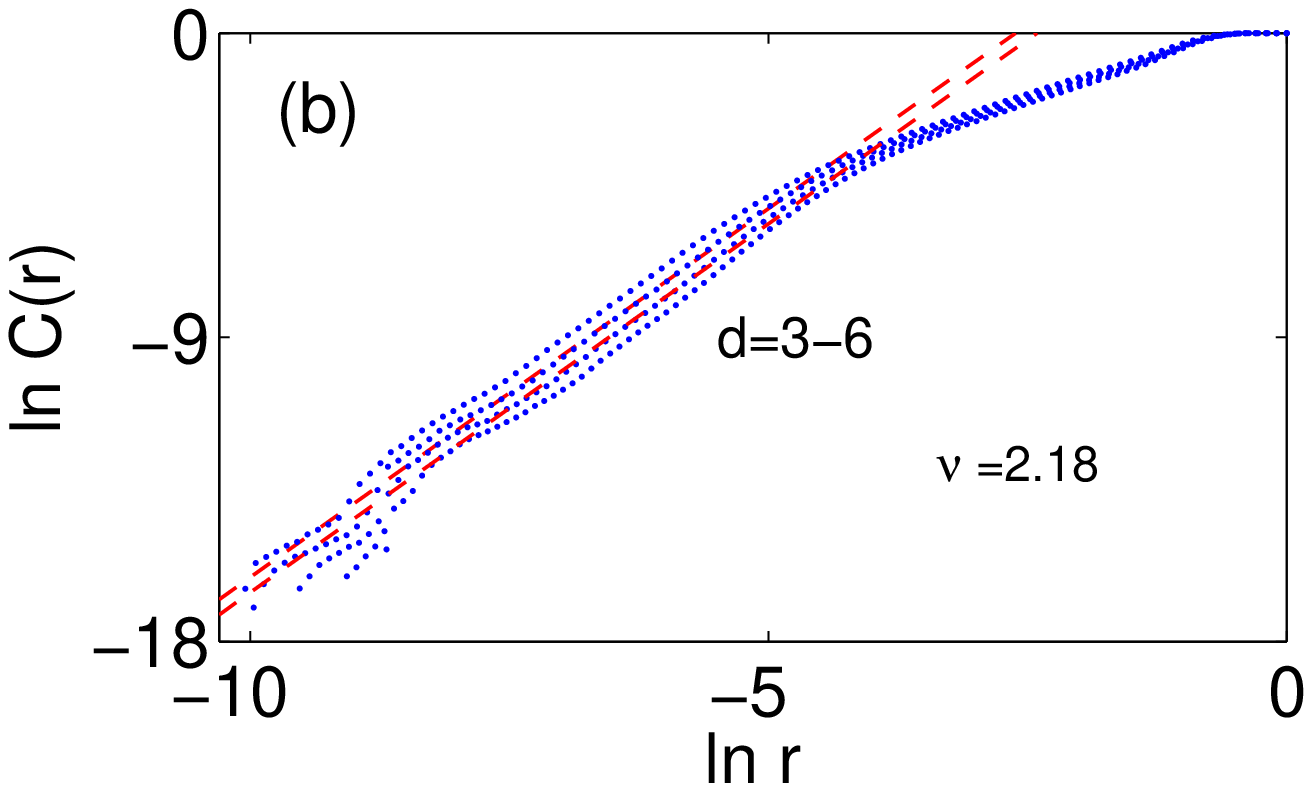}
}
\caption{(a and b)  Log-log plot of the correlation integral as a function of $r$ for the the first and the second half of the stress-strain series  for the less ductile $Zr_{65}Cu_{15}Ni_{10}Al_{10}$. The delay time is $12$. }
\label{Stest}
\end{figure}

We have calculated the correlation integral, which is known to be more sensitive to nonstationarity \cite{KS97} for the two halves of the data set and the full set. In general, using a much smaller delay time compared to the correlation time has been known to enhance the scaling regime. Thus, the delay time of $12$ sampling units of time is used. A window size of $w = 2(d-1) \tau$ is used to exclude temporally correlated points \cite{Theiler86}.    A log-log plot of the correlation integrals for the first and second half is shown in Fig. \ref{Stest} for $\tau=12$ and for embedding dimensions $d=3-6$. As can be seen, the  slopes for successive embedding dimensions converge to $\sim 2.2$ for both halves of the data set. This turns out, as we shall see, to be nearly the same as the converged value of the slope for the full data set. Thus the ''corrected'' data set is  taken to be stationary. 

\begin{figure}[h]
\vbox{
\includegraphics[height=5cm,width=8cm]{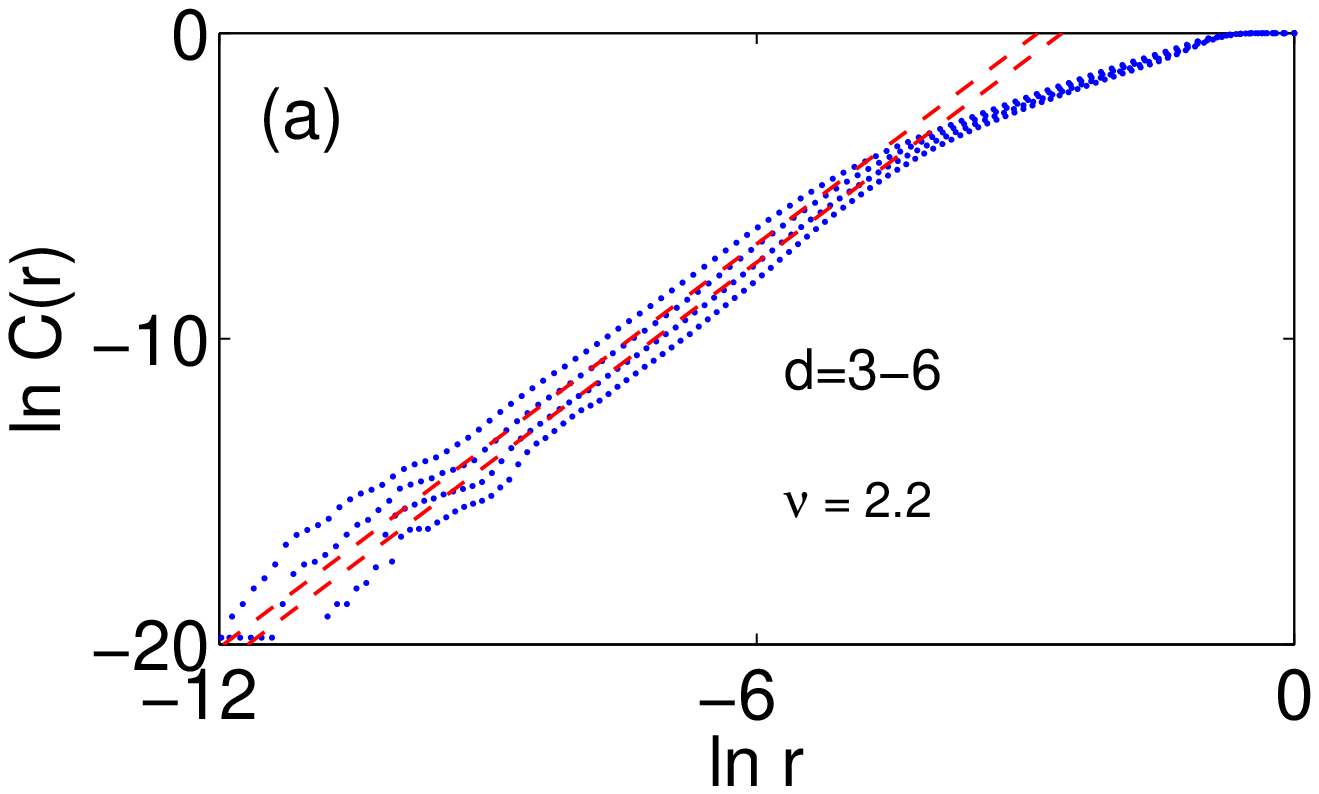}
\includegraphics[height=5cm,width=8cm]{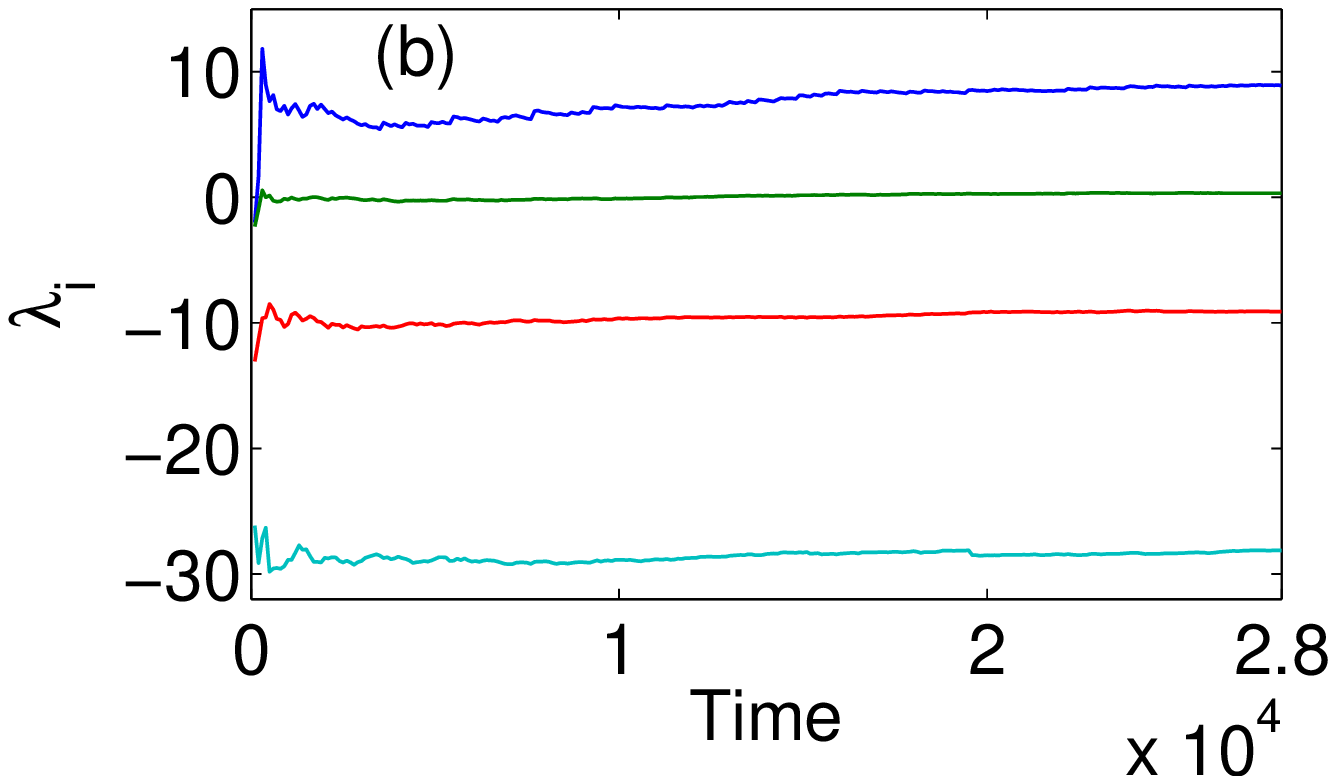}
}
\caption{(a)  Log-log plot of the correlation integral as a function of $r$ for the stress-strain series for the less ductile $Zr_{65}Cu_{15}Ni_{10}Al_{10}$ sample for $d=3-6$ with delay time $12$. (b) Plot of the Lyapunov spectrum for the same  time series with $d=4$ and delay time $2$.}
\label{Corrd1}
\end{figure}

To estimate the correlation dimension, we shall use the full data set. A plot of $ln \, C(r)$ verses $ln \, r$ for the full data set of $29,000$ points is shown in Fig. \ref{Corrd1}a for embedding dimensions $d=3-6$. It is clear that the slopes of $ln \, C(r) $ verses $ln \, r$ converge to about $2.2$ as the embedding dimension reaches $d=6$. Comparing these plots with those in Fig. \ref{Stest} (for the two halves of the data set), we see that the local slope fluctuations in Fig. \ref{Corrd1}a is much less than in Fig. \ref{Stest}. This feature is to be expected for a longer time series.  The converged value of the slopes of $\ln \, C(r)$ is taken to be the correlation dimension $\nu \sim 2.2$ of the signal. 

The Lyapunov spectrum is  calculated (for the full data set) keeping $d= 4$ with a time delay of $\tau=2$ and a shell size $0.3\%$. The spectrum is shown in Fig. \ref{Corrd1}b.  Note the existence of a stable positive and a zero Lyapunov exponent.  The Kaplan-Yorke dimension calculated from the Lyapunov spectrum turns out to be $D_{ky} \sim 2.84$, which is consistent with the correlation dimension.  The existence of a finite correlation dimension and a stable positive (and zero) Lyapunov exponent strongly suggests that the stress signal is chaotic. (We have also calculated the Lyapunov spectrum for the two halves of the data set. The values of the positive and zero exponents are close to those of the full data set.)

\begin{figure}[h]
\center{
\includegraphics[height=5cm,width=8cm]{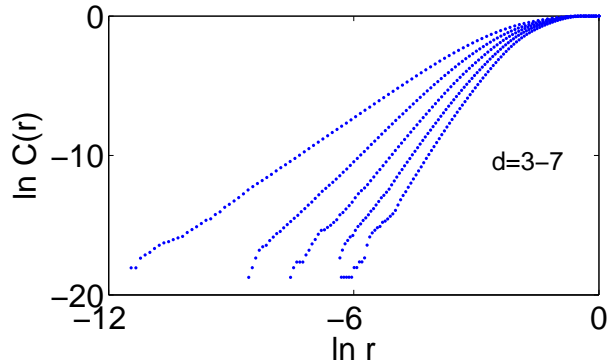}
}
\caption{(a)  Log-log plot of the correlation integral as a function of $r$ for the stress-strain series for the ductile $Cu_{47.5}Zr_{47.5}Al_{5}$ BMG sample for $d=3-7$ with delay time $12$. }
\label{Corrd2}
\end{figure}

Now consider the dynamic analysis of the time series for the ductile $Cu_{47.5}Zr_{47.5}Al_{5}$ BMG sample $S_D$.  The total length of the time series used is about $18,000$ points in units of $\delta t=0.05 s$, which is quite short. While the plastic strain is much larger than the less ductile sample $S_B$, the coarse sampling rate imposes limitation on the analysis. The auto-correlation time for the  stress-time series is about $18$ units of $\delta t$. In this case, systematic changes in the mean stress level have been removed by subtracting the running average and the amplitudes are rescaled using a local average. The resulting time series is expected to be stationary. The mean, variance of subintervals of the data set does not show much dispersion. The autocorrelation functions for the 
first and the second half of the time series remain nearly the same, giving the same correlation time as that for the full data set. This suggests that the time series is stationary. However, the stronger test for stationarity that should be reflected in the calculated correlation integrals from the first and  second half of the data set as with the full data set do not offer any additional support as the slopes of $ln \, C(r)$ verses $ln \, r$ do not converge even though the slopes for each embedding dimension  are nearly the same (not shown).  In view of this, we present the results for the full data set that shows no evidence of chaos. We have used a range of typical time delay values along with suitable window sizes to calculate the correlation integral.  However, we have not been able to 
find any convergence in the slopes of the correlation integral.  A plot of $ln \, C(r)$ verses $ln \, r$ is shown in Fig. \ref{Corrd2} for embedding dimensions $d=3-7$. It is clear that there is no evidence of  convergence of the slopes even for $d=7$. A calculation of Lyapunov spectrum also did not show the existence of positive exponent or a zero exponent. On the basis of this we conclude that there is no evidence of chaotic dynamics.

\subsection{Self-organized critical dynamics}

\begin{figure}[h]
\vbox{
\includegraphics[height=5.0cm,width=8cm]{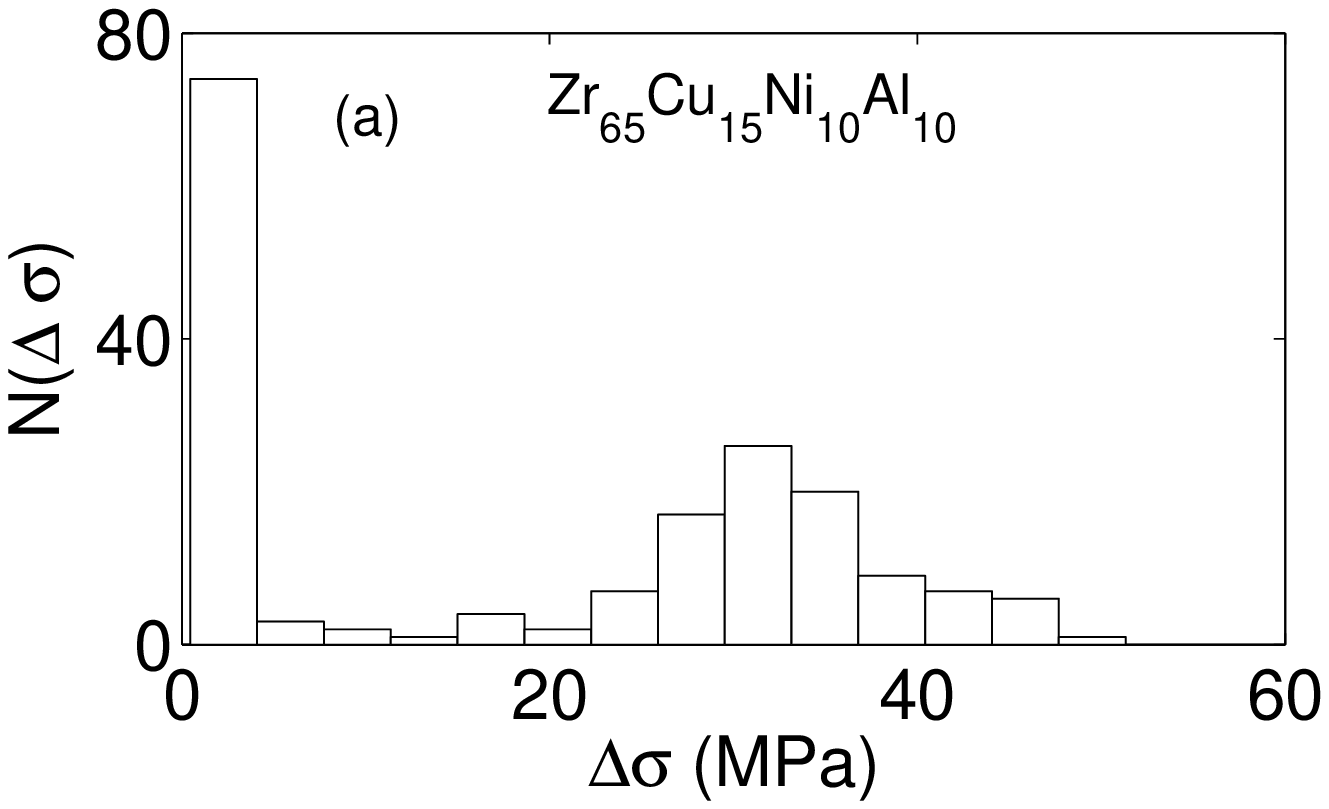}
\includegraphics[height=5.0cm,width=8cm]{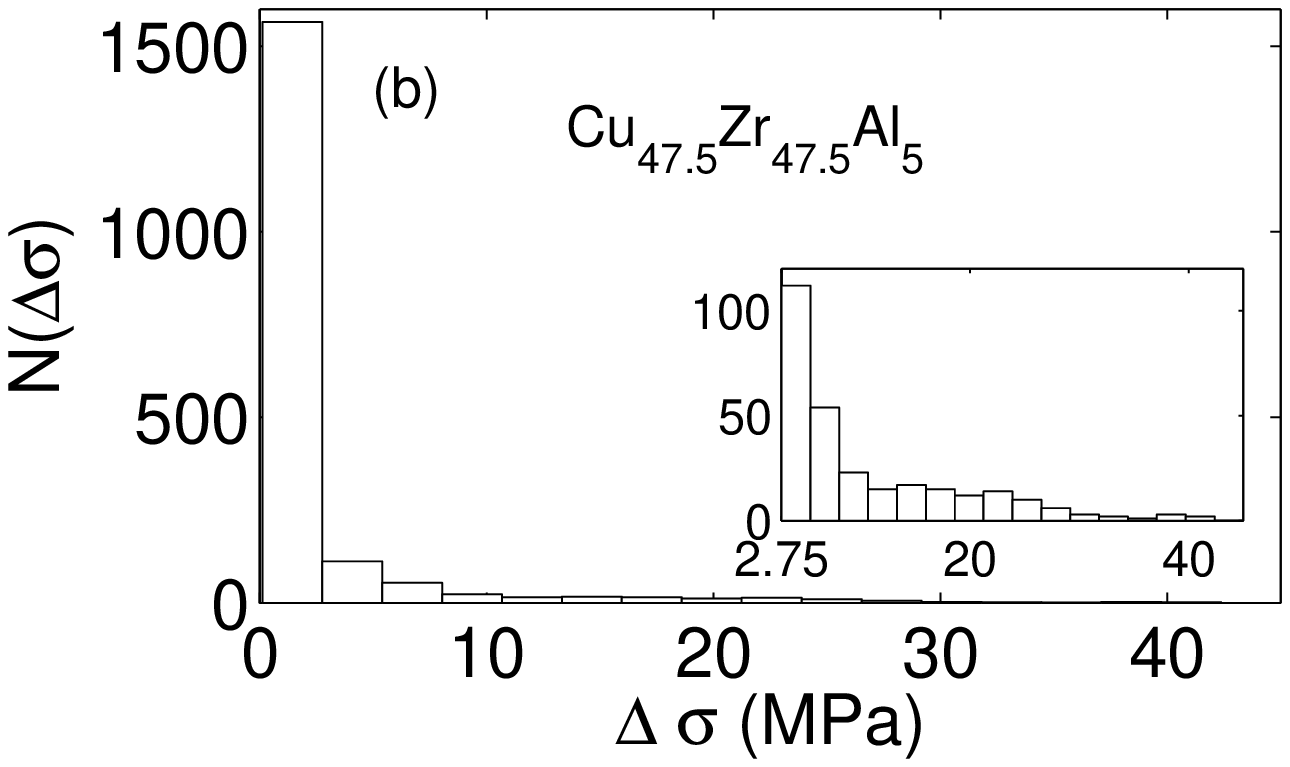}
}
\caption{(a and b) Distribution of stress drop magnitudes for the less ductile $Zr_{65}Cu_{15}Ni_{10}Al_{10}$ and more ductile $Cu_{47.5}Zr_{47.5}Al_{5}$ BMG samples, respectively.}
\label{Distribution}
\end{figure}

A statistical analysis of the stress-strain curves for both the samples has also been been carried out. First, consider the less ductile sample $S_B$. By visual inspection, it is clear that most stress drops are large with much fewer small stress drops. This is what is reflected in the distribution of stress drop magnitude shown in Fig. \ref{Distribution}a. The figure shows a nearly symmetric distribution around an average value of $30 MPa$, but there is a peak for small values (about $2 MPa$). The corresponding distribution of the durations also exhibits a similar form. No scaling regime could be detected in these distributions.

\begin{figure}[h]
\vbox{
\includegraphics[height=5.0cm,width=8cm]{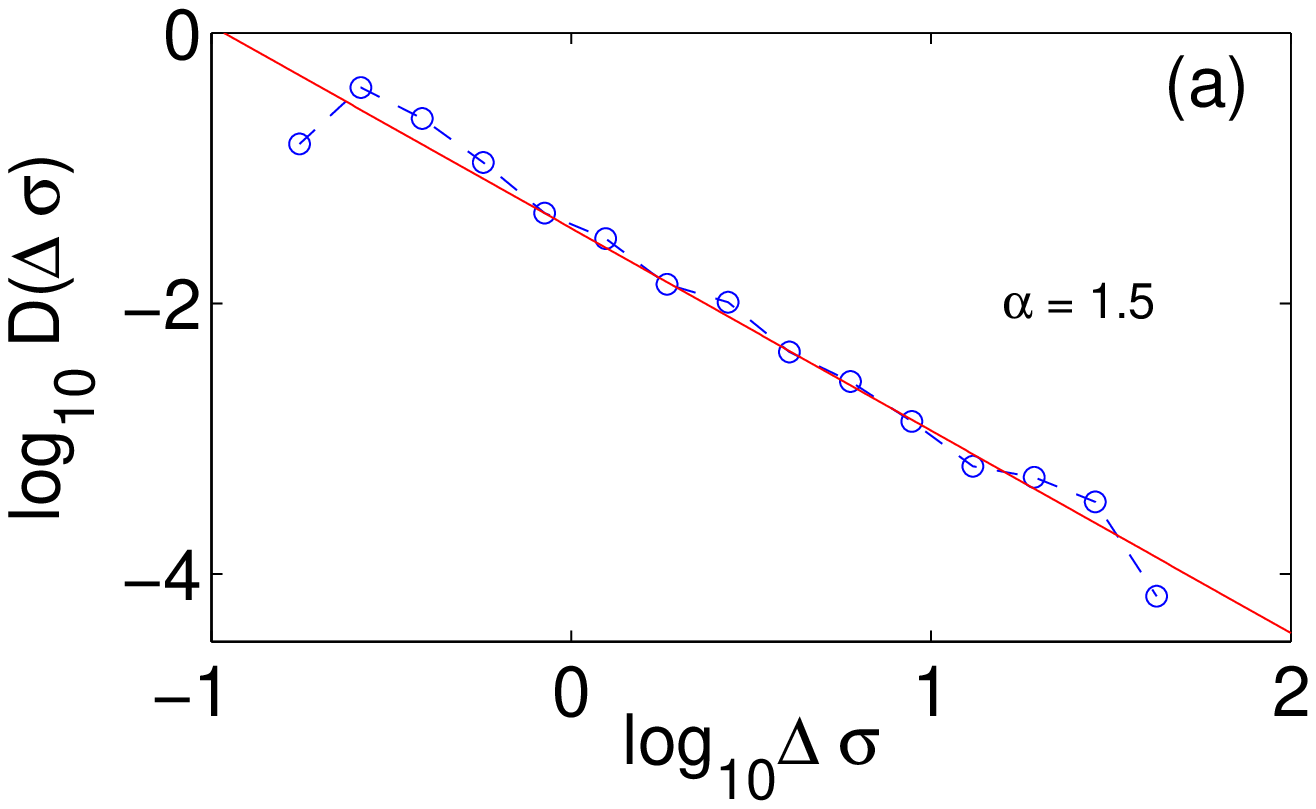}
\includegraphics[height=5.0cm,width=8cm]{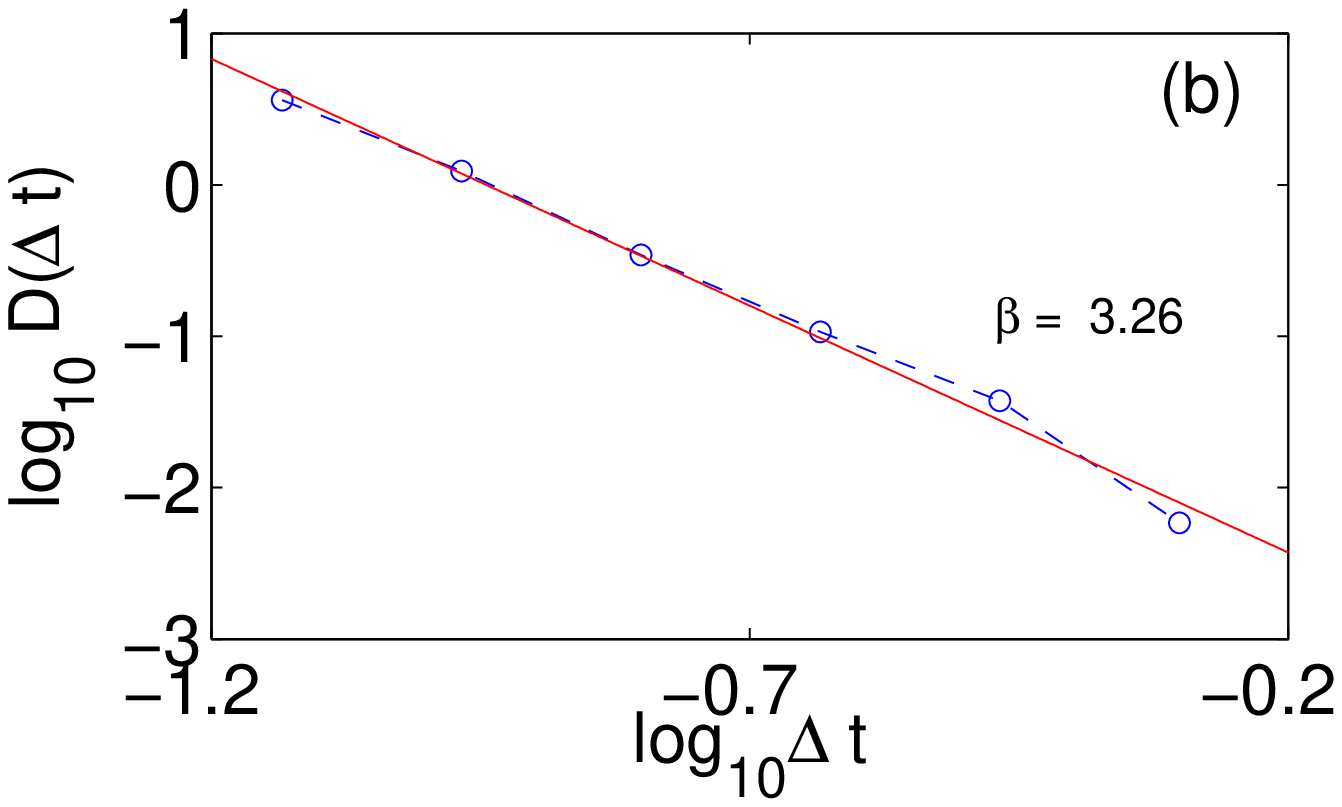}

}
\caption{(a and b) Distribution of stress drop magnitudes and durations of the ductile $Cu_{47.5}Zr_{47.5}Al_{5}$ sample, respectively.}
\label{Dstress2}
\end{figure}

In contrast to the less ductile sample $S_B$, a visual inspection of the stress-time series for the ductile sample $S_D$ suggests that there are a large number of small stress drops with fewer large drops. This feature is similar to the high strain-rate type A serrations of the PLC effect \cite{Bhar01,GA99} wherein the analysis showed a power-law distribution of stress drops. Following this, we have calculated the  distribution of the stress drops as a function of the magnitude of  the stress drops. As can be seen from Fig. \ref{Distribution}b, the distribution of stress drops shows a monotonically decreasing trend, which when plotted on a log-log scale shows that the  distribution follows a  power law of the form $D(\Delta \sigma)\sim \Delta \sigma^{-\alpha}$.  This is shown in Fig. \ref{Dstress2}a. The scaling regime is nearly two  orders in $\Delta \sigma$  with an exponent value $\alpha \sim 1.5$. 

\begin{figure}[h]
\center
\includegraphics[height=5cm,width=7.5cm]{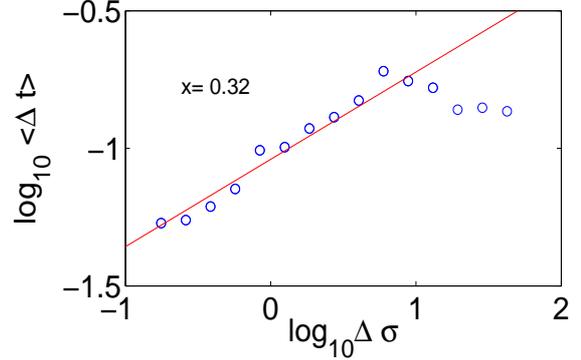}
\caption{Mean duration of the stress drops  as a function of the size of the drops for $Cu_{47.5}Zr_{47.5}Al_{5}$ sample. }
\label{TS2}
\end{figure}

Since the stress drop magnitudes are related to their durations, a similar analysis for the distribution of the durations of the stress drops $\Delta t$ can be investigated for the ductile case. The distribution shows a tendency to follow  a power-law  scaling of the form $D(\Delta t)\sim \Delta t^{-\beta}$, as reflected in Fig. \ref{Dstress2}b. The exponent $\beta \sim3.26$. However, the extent of scaling regime is much less than that for the stress drop magnitudes.  Such a limited scaling regime in the event durations is well documented. Indeed,  even in model systems, such as the sand-pile model \cite{Bak88,Manna90}, the extent of scaling regime in durations is much less than that for the magnitudes. \footnote{For instance, in the large-scale simulations of sand-pile model \cite{Manna90}, 
while the scaling regime  for the magnitudes is nearly five order, that of durations is at best three orders. Further, in most  experimental work that we are aware of where scaling for magnitudes and durations are reported, the scaling for duration is seldom more than one and half orders \cite{Bak96}. Quite often only the distribution of event sizes is taken to identify the process as SOC \cite{Turcotte99}.}. For model systems, it is easy to trace this features to the fact that the life time of a cluster is much shorter than its size \cite{Bak88}. In this case, since $\beta$ (the exponent for durations) is high, sampling rate should be high for the same scaling extent to be realized compared to a situation where $\beta \sim 1$ as in the case of the PLC effect. In addition, in our case, part of contribution to limited scaling regime arises due to the coarse sampling rate ($20Hz$) and  due to the limited precision in the measurement of the stress.   We have also plotted the mean durations of the stress drops as a function of  the size of the stress drops shown in Fig. \ref{TS2}.  The exponent value is $x \sim 0.32$. The scaling regime is nearly two orders as for the stress drop magnitudes. It can be easily checked that the scaling relation $\alpha = x(\beta -1) + 1$ is  satisfied reasonably well. 

An independent check of the scaling relation can be carried out by computing the exponent corresponding to the power spectrum of the stress-strain curve. Kert\'{e}sz and Kiss \cite{KK90} derived  an alternate relation between the exponents $\alpha$ and $x$ that relates to the low-frequency behavior of the power spectrum.  By assuming that the total energy dissipated stems from independent elementary events whose energy density spectrum is quasi-Lorentzian, these authors have shown that if the scaling exponent satisfy the inequality
\begin{equation}
2x+\alpha > 3,
\end{equation}
then the lower-frequency component of the power spectrum $S(\omega)$ should behave as:
\begin{equation}
S(\omega) \sim  \omega^{-(3-\alpha)/x}.
\end{equation}
Thus, if $2x+\alpha<3$, the $S(\omega)$ will always scale with $1/\omega^{2}$.

\begin{figure}[h]
\hbox{
\includegraphics[height=5cm,width=8.0cm]{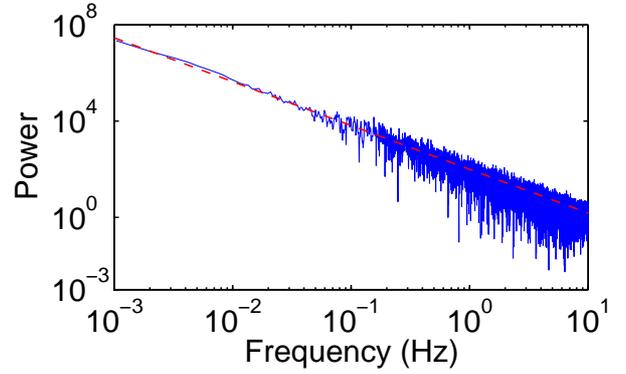}
}
\caption{ Power spectrum for the first half and the full data of the uncleaned stress-strain series respectively of the ductile $Cu_{47.5}Zr_{47.5}Al_{5}$ BMG 
sample. }
\label{Power}
\end{figure}

The power spectrum of the time series for the ductile alloy for the two halves and the full data set have been calculated.  The low-frequency region for these three data sets can be well fitted to a power law with nearly the same exponent value $\sim 1.9$. Since the plots are practically the same, in Fig.  \ref{Power} we show the power spectrum for the full data set. This is clearly consistent with the expected slope of  $1/\omega^{2}$ as $2x+\alpha<3$.  

Despite the limited scaling regime in the time durations, the cross check increases the confidence level in asserting the SOC nature of the stress-time series of the ductile sample.

Finally, some comments may be useful on issues of nonstationarity. We note that the stress-time series of the ductile sample shows hardening, while that of the less ductile sample is nearly constant. However, the ductile data, which requires subtracting the running average from the raw data set and rescaling, is not chaotic, but displays power law distribution. However,  the power-law nature of the distributions remains the same for the detrended data set or the raw data set as the definition of the event is not influenced by drift of the mean stress level. Thus, the results of the analysis are not affected by subtracting the mean drift in stress-time series.  In contrast, the less ductile sample data is nearly flat, but is chaotic, requiring stationary time series. In this sense, removing the 
mean stress level from the time series has very little effect on our results.

\section{Summary and discussion}
In summary, the analysis of the stress-strain curves from the intermittent flow of the two BMG samples shows that the deformation dynamics is very complex. Two complex but distinct dynamical regimes, namely chaos and SOC, are discovered.  The underlying dynamics responsible for irregular serrations in the less ductile $Zr_ {65} Cu_ {15} Ni_ {10} Al_ {10} $ BMG is chaotic; chaos requires a small number of collective degrees of freedom for a proper description of the dynamics.  Indeed,    since the associated correlation dimension is $ \nu \sim 2.2$, the minimum number of degrees of freedom, i.e., the minimum number of variables required to describe the dynamics of serrations is given by the integer larger than $\nu + 1 $, which in this case is four \cite{Ding93}. In contrast, for the ductile $Cu_{47.5}Zr_{47.5}Al_{5}$ BMG, power law distributions are seen for stress drop magnitudes and durations. While two orders of magnitude for the extent of scaling for the stress drop magnitudes (and the mean durations of stress drops) is reasonable for any experimental data, the extent of scaling for durations is only one order.  Considering the coarse sampling rate as well as the limited control in measuring the stress, coupled with the fact that the exponents also satisfy the scaling relation and the independent check using power spectrum, it appears that  the ductile $Cu_{47.5}Zr_{47.5}Al_{5}$ BMG follows  SOC type of dynamics. Due to the scale-invariant nature of the process,  SOC dynamics requires large number of degrees of freedom (actually infinite in theory), which in turn suggests that a hierarchy 
of length scales is responsible for the stress-strain curves. Noting that stress-strain curves  are  a result of spatial average over all the shear band activity in the entire sample, the fact that {\it two different kinds of dynamics could be detected comes as a surprise.} Nevertheless, the nature of the serrations appears to contain this hidden information. The next step is to interpret these results in the light of available physical information and models to see if further insight is possible. In the following we attempt to build a physical picture that is consistent with the present state of knowledge.

To this end, we note that there is considerable similarity between the serrated flow of the BMGs and the PLC effect even though the underlying physical processes are different.  Thus,  much can be inferred from the similarities and the differences between the serrated flow in the BMGs and the PLC effect \cite{Bhar01,Bhar02a,GA99,Noro97}. Recall that  the PLC  serrations and the associated bands arise due to collective pinning and unpinning of dislocations from solute atmosphere, corresponding to the stick and the slip phases, respectively. Thus, longer the waiting time, larger is the stress required to unpin them. For low strain rates, there is adequate time for aging to be completed, which therefore results in large yield drops. Increase in applied strain rate restricts the aging process, leading to a decrease in the serration amplitudes \cite{GA07,Kubin87}.

In the case of the PLC effect, at low strains, large amplitude type C serrations are seen \cite{GA07,Kubin87}. Time-series analyses of the PLC type C serrations (in different sample types) have revealed that the type C serrations are chaotic \cite{Bhar01,Bhar02a,Noro97}.   On the other hand, the type C serrations  are identified with randomly nucleating dislocation bands {\it where each stress drop can be identified with a single band, i.e.,  the entire plastic strain rate is localized  to a small spatial extent} \cite{GA07,Kubin87}. This picture is substantiated by the Ananthakrishna model for the PLC effect \cite{GA07,Anan04,Bhar03a}. The similarity with the large-amplitude chaotic serrations of the  $Zr_{65}Cu_{15}Ni_{10}Al_{10}$ BMG sample is clear.  The  fact that the PLC type C serrations are 
due to localization of dislocations into bands {\it suggests that each stress drop in the case of the $Zr_{65}Cu_{15}Ni_{10}Al_{10}$  BMG sample  may be a result of localization of the shear to a small spatial region.} This inference appears to be supported by recent observations on samples that exhibit significant ductility despite a single shear operation. In such ductile BMGs, each stick-slip event has been shown to correspond to localized striations (of regular spacing)  \cite{Cheng09,Song08,Han09}. These studies attribute the significant ductility to repetitive sliding of a major shear band.  In contrast, a vein-like structure is usually found in low ductility situations where catastrophic failure is triggered by a single shear band.  

Indeed, Cheng {\it et al.,} have proposed a stick-slip model  that is applicable to the instability process when a major shear band is observed \cite{Cheng09}. These authors show that in this case the temperature rise is small, with the deformed sample remaining  cold. While it is not easy to measure the in situ temperature inside the sample, an estimate based on their model turns out to be small for our case. Noting that the composition of our sample is almost the same as used in their experiment, it appears that their model adequately captures the results of the dynamics of serrations in the less ductile BMG.

In general, stick-slip dynamics is considered to be regular, meaning that they can be approximated by periodic recurrence of stress drops (see Ananthakrishna and De in Ref. \cite{BK06}.) The present analysis, however, shows that the serrations are not regular but chaotic with minimum of four variables required for dynamic description of serrated flow \cite {Ding93}. Identifying these four degrees of freedom with physical variables should give some insight into the underlying dynamics.  This will be done based on known physical mechanisms and model studies used in describing the deformation process. 

Several theoretical studies has demonstrated the importance of free volume and local temperature in understanding the deformation of BMGs 
\cite{Spaepen77,Spaepen82,Argon79,Falk98,Langer01,Cheng09}. It is well known that  shear  localization is affected by rise in the local temperature and  volume dilation due to free volume generation. The local plastic strain rate depends on the available free volume, temperature and the local stress within the shear band.  The latter is controlled by applied strain rate through the machine equation (stress rate is proportional to the difference between the applied and plastic strain rate). Thus, the three obvious variables that can be identified are the local plastic strain rate, free volume and temperature.  However, both temperature and free volume in the shear band are influenced by the applied  strain rate, which also affects the instantaneous flow stress in the shear band. Moreover, since the applied strain rate limits the extent of contribution from these three variables, which in turn depends on stress, the instantaneous stress in the shear band can be identified as the fourth variable. Thus, we rationalize that  a minimal model should include local plastic strain rate, free volume, temperature and stress. Indeed, these are the four variables used in the model proposed by Cheng {\it et al.} \cite{Cheng09}. However,  if the intended goal is to model the fluctuating stress-strain curves as the spatial average over the shear band activity in the entire sample, this objective appears to be beyond the current state of knowledge.

In the case of the ductile $Cu_{47.5}Zr_{47.5}Al_{5}$ BMG, the serrations are quite irregular with large number of small stress drops.  This feature is similar to the PLC type A serrations seen at high strain rates for the  propagating bands \cite{GA07,Anan04,Bhar03a,Bhar01,Bhar02a,Noro97}.  The small magnitude stress drops are attributed to incomplete plastic relaxation.  In the PLC type A serrations, power-law distributions of stress drop magnitudes and durations have been reported \cite{Bhar01,Bhar02a,GA99}. Since the serrations from the ductile BMG also have a large number of small stress drops as in the PLC type A serrations,  the possibility of a power-law SOC state is high.  While the extent of the scaling is somewhat limited, an SOC type of dynamics is suggested.  We note here that even in the case of PLC type A serrations, the sampling rate is not significantly better but the control on stress was improved.  Given these limitations, the interpretation of the physical origin of the SOC-type dynamics could be somewhat speculative.

To understand the origin of the power-law distribution of stress drop magnitudes and durations in the ductile $Cu_{47.5}Zr_{47.5}Al_{5}$ BMG sample,  consider Fig. \ref{Bands}b. In this case,  multiple shear bands could be observed on the surface of the deformed sample. Further, the magnified view of a severely deformed region shown in  Fig. \ref{Bands}c clearly suggests a complex interaction among the large number of shear bands. This also suggests that the possibility of a hierarchy of length scales over which the shear can propagate. Thus, a large number of metastable states can be created due to the interaction between the shear bands. This includes short-range interaction from the intersection of the bands and consequent arrest, and long-range interaction through the strain fields.   Then, each stress drop corresponds to the system surmounting the barrier (corresponding to the arrested shear bands) jumping to a neighboring metastable state only to be caught in another metastable state.  This picture is consistent with SOC-type dynamics. 

Liu {\it et al.} \cite{Liu07} have shown that the large plasticity is caused by the simultaneous nucleation of numerous shear bands (in different directions) and their interactions.  The microstructure of the ductile $Cu_{47.5}Zr_{47.5}Al_{5}$ BMG has been shown to consist of hard regions surrounded by soft regions. The wavelength of the wavy shear bands is consistent with the dimensions of the hard regions. They ascribe the plasticity of the ductile BMGs to concurrent nucleation and hindered propagation of the shear bands throughout the sample. From this point of view, SOC-type of dynamics is likely to be seen in many other ductile metallic glasses \cite{Johnson04,Yu08,Liu07,Gu06,He03,Xing01,Das05,Inoue05} \footnote{ The basic idea used in designing ductile BMGs is to nucleate large number of shear bands that can interact with each other, thereby preventing a catastrophic failure. This can be accomplished, for example, by providing a heterogeneous microstructure such as nanoscale crystallinity \cite{He03,Xing01,Das05,Inoue05}.}.  Indeed, Wang {\it et al} \cite{Wang09} show that there is a scaling regime for the distribution of elastic energy density  for several samples that are relatively more ductile.  
 
However, one basic difference between the stress-strain curves of the ductile BMG and the PLC type A serrations is the sharpness of the stress drops in the former case as can be seen from Fig. \ref{OrigSScurve1}b. Indeed, the reloading part is nearly linear except for a possible small curvature seen just before the stress drop. This must be contrasted with the stress-strain curves for the type A serrations of the PLC effect, where there is a substantial deviation from linearity, suggesting a plastic contribution on the loading part. More importantly, there are small stress fluctuations both on the ascending part as well as the descending part of the stress-time curves of the PLC type A serrations. (See Fig. 1e of Ref. \cite{Bhar01}.) {\it The latter feature is the signature of partial relaxation 
of plastic strain bursts.} Thus, the sharp stress drop for the ductile BMG also implies that all shear band activity is close to the fully  relaxed state. 

Finally, we note that while the less ductile sample is chaotic it does not follow SOC dynamics (i.e. the stress drop magnitudes and durations do not follow power-law distribution).  In contrast, the ductile sample that follows SOC dynamics does not show any evidence of chaos. This is consistent with the well-established fact that these two dynamic states are considered to be very distinct:  chaos is characterized by a few degrees of freedom, while SOC is characterized by a large number of degrees (theoretically infinite) of freedom. Due to this fundamental distinction, these two types of dynamics are considered mutually exclusive and are not realized in the same system. Indeed, in the PLC effect also, the chaotic type C and B serrations are observed at low and medium strain rates, while type A serrations that follows SOC dynamics are observed at high strain rates \cite{Bhar01,Bhar02a,GA99}. 

In summary, we have analyzed the serrated flow behavior for the two typical BMGs by time-series analysis and statistical methods. For the less ductile alloy, we have demonstrated the chaotic nature of the stress serrations by showing the existence of a finite correlation dimension and a  positive Lyapunov exponent. In contrast, for the ductile alloy, the distributions of stress drop magnitudes and their time durations appear to follow power-law scaling form reminiscent of self-organized critical state. The three exponents also satisfy the scaling relation that should be satisfied by power-law SOC dynamics \cite{Bak88,Jensen98}.    {\it To the best of our knowledge, this is the first time that the two dynamical regimes are observed in the deformation of the two different  BMGs.} Considering the fact that the  extent of plasticity is only  a factor of two between the two samples, and the fact that the method of analysis uses only the stress-strain curves,  {\it this difference in the nature of dynamics is surprising.}  Indeed, that the method is able to detect this difference is {\it also} surprising. However, we have suggested that the basic difference in the dynamics can be related to the differences in the physical mechanisms contributing to the different ductility levels in the two BMGs. The chaotic dynamics of the less ductile BMG appears to be consistent with the accepted picture of a single shear band sliding in repetitive manner, leading to considerable ductility. Each stick-slip corresponds to localized striations of regular spacing in samples \cite{Cheng09,Song08,Han09}. Further, for this case, the minimum number of variables required for the description of the chaotic serrations is four. On the basis of known physical mechanisms,  we have identified these variables to be the free volume, temperature, local plastic strain rate and stress; this is consistent with the model of Cheng {\it et al} \cite{Cheng09}. In contrast, for the ductile BMG, the high ductility appears to be related to concurrent 
nucleation of large number of shear bands throughout the sample, which in turn can give raise to a hierarchy of length scales over which shear bands propagate.  Our analysis also suggests that the ductile nature emerges from the production of several shear bands and their interactions rather than partial plastic relaxation of the shear activity in the shear bands as in the PLC effect.  These rationalizations  may have implication in understanding the mechanism controlling plastic deformation of the ductile BMGs and should prove useful in modeling efforts.

\centerline{
{\bf ACKNOWLEDGMENTS}}
GA would like to acknowledge the Department of Atomic Energy grant through the Raja Ramanna Fellowship scheme and also support from BRNS Grant No. $2007/36/62$-$BRNS/2564$. RS would like to acknowledge C. S. I. R. India for financial support. W. H. Wang would like to acknowledge the support from the NSF of China (No. $50731008$ and No. $50921091$) and Most $973$ (No. $2010CB731603$).





\bibliographystyle{model1-num-names}
\bibliography{<your-bib-database>}



\end{document}